\documentclass[twocolumn,aps,prl,epsf,graphics,psfig]{revtex4}
\usepackage{graphicx}

\begin{document}
\title{
Spin-blockade effect and coherent control of DNA-damage by free radicals:
a proposal on bio-spintronics
}

\author{Ramin M. Abolfath$^{1,2}$, Thomas Brabec$^1$}

\affiliation{
$^1$Physics Department, University of Ottawa, Ottawa, ON, K1N 6N5, Canada \\
$^2$School of Natural Sciences and Mathematics, University of Texas at Dallas, Richardson, TX 75080
}

\date{\today}

\begin{abstract}
Coherent control of OH-free radicals interacting with the spin-triplet state of a DNA molecule is investigated.
A model Hamiltonian for molecular spin singlet-triplet resonance is developed. We illustrate that the spin-triplet state in DNA molecules can be efficiently populated, as the spin-injection rate can be tuned to be orders of magnitudes greater than the decay rate due to small spin-orbit coupling in organic molecules.
Owing to the nano-second life-time of OH free radicals, a non-equilibrium free energy barrier induced by the injected spin triplet state that lasts approximately longer than one-micro second in room temperature can efficiently block the initial Hydrogen abstraction and DNA damage.
For a direct demonstration of the spin-blockade effect, a molecular simulation based on an {\em ab-initio} Car-Parrinello molecular dynamics is deployed.
\end{abstract}
\pacs{87.56.-v,87.55.-x,87.64.Aa}
\maketitle

Organic molecules including DNA have been proposed to be used in electronic circuits as molecular conductors and nanowires~\cite{Chakraborty:book}. They are attractive alternatives to inorganic semiconductors for a variety of optoelectronic devices, e.g. photovoltaics. Their long spin lifetime~\cite{Gorner1990:JPP} can make them desirable for applications in spintronics~\cite{Zutic2004:RMP}.
The unique biological functionality of DNA-molecules in carrying the genetic information can be used for {\em bio-spintronics}.

We investigate here the possibility to couple coherent control methods \cite{Rabitz2000:S} with bio-spintronics.
OH-free radicals that are products of chemical reactions or ionizing radiation~\cite{EricHall:book} are highly reactive and toxic,
resulting in DNA-damage (single/double strand break), and the development of genetic aberrations, carcinogenesis and aging.
They are highly unstable diatomic molecules with a life-time that is reported within nano-seconds and are charged neutral with a magnetic moment that is produced by nine electrons with an unpaired spin in the outermost open shell.
Their high reactivity is thus attributed to the pairing of opposite spin electrons in the open orbital of the free radical with an
electron in an organic molecule. Fenton chemistry studies used in nucleic acid foot-printing show that a free radical reaches a DNA-molecule by diffusion and removes a hydrogen ion from its sugar moiety (e.g., a deoxyribose)
~\cite{Balasubramanian1998:PNAS,Tullius2005:COCB,Pogozelski1998:CR}.
The most probable sites of attack are identified as H$_{5'}$, H$_{5''}$, and H$_{4'}$ of the deoxyribose residue in the backbone of the DNA (the nomenclature is adopted following Ref.~\cite{Pogozelski1998:CR}).
Because of their short life-time of 1 ns, OH radicals generated within 1 nm from the surface of the DNA molecule can remove a hydrogen ion and form a water molecule.

We show here that by optical pumping the HOMO electron of both, the OH-free radicals and the DNA, can be efficiently spin-polarized.
Paramagnetic state of OH-radical can be oriented along the non-equilibrium direction of the triplet magnetic moment of the DNA molecule.
Such photo-excited DNA follow a phosphorescence decay to the ground state with slow transition rate, hence providing statistically enough triplet excitons to block OH-DNA reactivity.
The reported triplet life-time of DNA nucleotides ranges 1$\mu$s at room temperature
to 1s at 77K~\cite{Gorner1990:JPP}.
The Hydrogen abstraction occurs in 1 ps, several order of magnitudes faster than the DNA triplet decay process.  The disparity in time scale between the chemical reaction and DNA-triplet life-time allow us to use a ps Car-Parrinello {\em ab-initio} molecular dynamics to simulate the immunization of DNA against OH-free radicals.

\begin{figure}
\begin{center}
\includegraphics[width=1.0\linewidth]{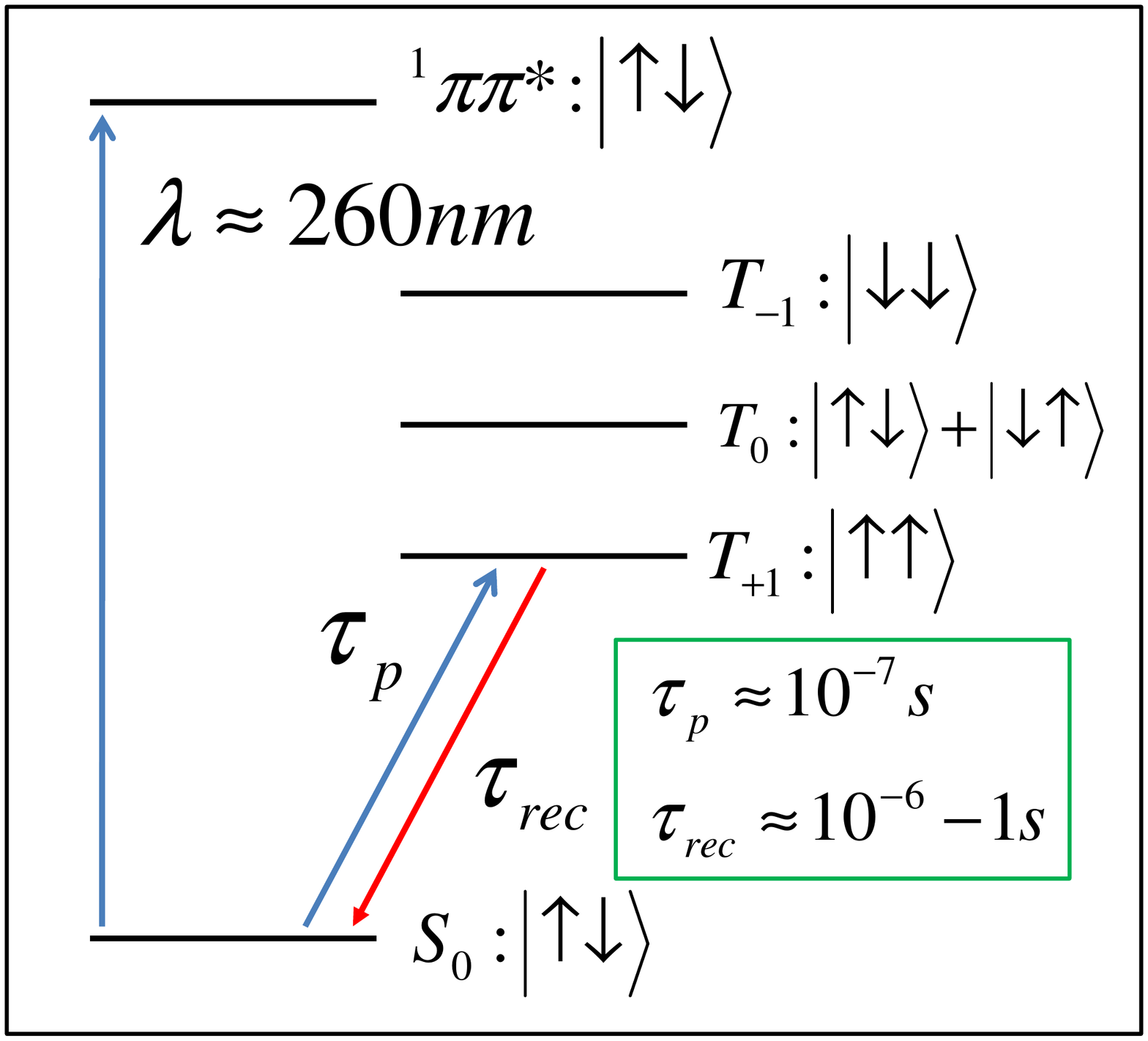}\vspace{-0.5cm} \\
\includegraphics[width=1.0\linewidth]{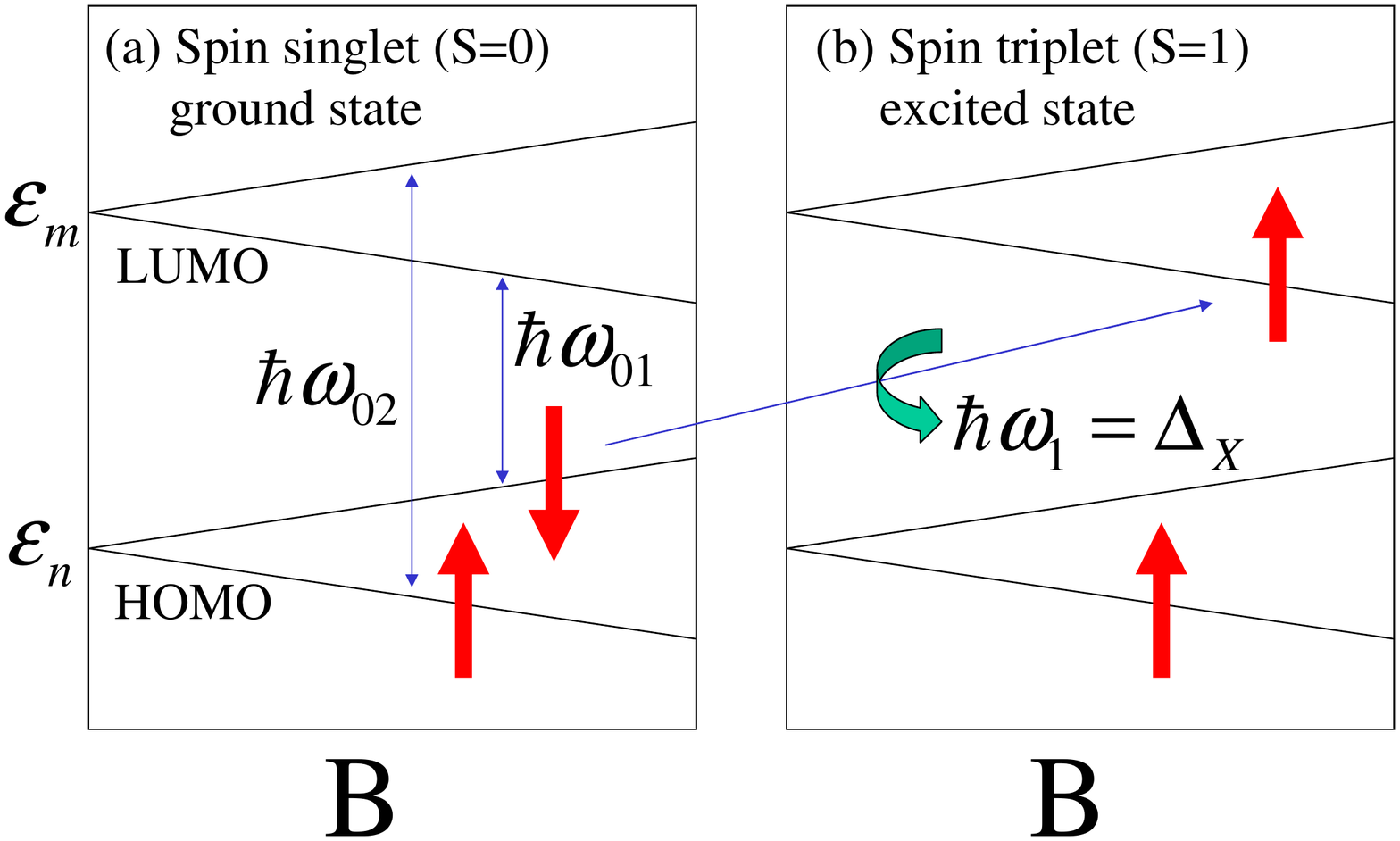}\vspace{-0.5cm}
\noindent
\caption{
Schematically shows the typical energy level diagram (top)
and the direct transition of spin singlet-triplet resonance (bottom) in the presence of external magnetic field.
}
\label{Fig0}
\end{center}
\end{figure}

Recent studies in pump and probe femto-second laser spectroscopy of nucleic-acids in solutions have provided valuable information on the optical excitations of nucleotides and their decay pathways to the ground state through radiative and non-radiative channels~\cite{Middleton2009:ARPC,Serrano2007:JPCB,Mercha2005:JACS,Hare2006:JPCB}. These studies have identified the spin multiplicities of the excited states including their life-times and quantum yields. As shown in Fig.~\ref{Fig0} the major photo-excitation of all bases of nucleic-acids occur close to the wavelength $\lambda=260$ nm (UV) where a direct transition from singlet ground state $S_0$ to singlet excited state $^1\pi\pi^*$ is observed.
Unlike III-V/II-VI semiconductors that strong SO-coupling has led to direct observation of the optical spin excitations in which a laser generates a population of photo-carriers strongly spin polarized along a direction which depends on the polarization state of the laser field~\cite{optical_orientation}, in organic molecules no direct optical transition to spin triplet state
has been reported.
However, the emission spectrum governed by inter-system crossing and SO coupling
has revealed the spin triplet states, $^3\pi\pi^*$, with energy below $^1\pi\pi^*$.
Consistent with the emission spectrum, our ab-initio calculation shows that
the lowest energy singlet-triplet energy gap ($\Delta_X$) of
DNA-bases and DNA-backbone are close to 2.68 eV (corresponding to $\lambda=463$ nm) for
DNA-base, and 2.38 eV ($\lambda=521$ nm)
for sugar moiety.

We employ a model Hamiltonian consisting of four HOMO-LUMO molecular levels as shown in Figure~\ref{Fig0}. 
The figure shows schematically the direct photo-induced spin singlet-triplet transitions ($S=0 \rightarrow 1$).
The molecular levels shown in Figure~\ref{Fig0} represent the ground and excited states of the DNA-molecule.
In the absence of an external magnetic field, the electronic ground state of the DNA-molecule is $S=0$ spin-singlet with double occupancy of the molecular orbitals.
The two-fold degeneracy of the molecular orbitals is lifted by
a weak magnetic field that couples to the electron spin through the
Zeeman interaction \cite{LandauLifshitzQM:book},
$E_{Z}= g\mu_B \vec{S}\cdot \vec{B}$.
Here $E_{Z}$ is the Zeeman energy,
$g$ is the electron $g$-factor ($g\approx 2$),
$\mu_B$ is the Bohr magneton ($\mu_B\approx 5.8\times 10^{-5}$ eV/Tesla),
and $B$ is the strength of the external magnetic field along quantization
axis, $\hat{z}$.

In the Born-Oppenheimer approximation, the $S=1$
spin-flip excitation is constructed by removing an electron from an occupied
$S_0$ state and putting into an unoccupied state with opposite spin.
By denoting quasi-particle energy levels (electron dressed by interaction)
by $\varepsilon_{n\sigma} = \epsilon_n + \Sigma(n, \sigma)$,
the energy of one exciton is
$\Delta_X = \varepsilon_{m\uparrow} -
\varepsilon_{n\downarrow} -
\langle m\uparrow,n\downarrow|V|n\downarrow,m\uparrow\rangle$.
Here the indices n and m label HOMO and LUMO, and
$\sigma= +1,-1$ is used for spin up and down.
$\epsilon_n$ is the nth single particle molecular energy level,
$\Sigma(n, \sigma) = \sum_{n'\in\rm occ.} [2\langle n,n'|V|n',n\rangle
- \langle n,n'|V|n,n'\rangle]$ is the electron self-energy and summation has
carried out over all occupied orbitals,
$V$ is the electron-electron repulsive Coulomb interaction,
and $\Delta_X = E_{\rm triplet} - E_{\rm singlet}$
is the energy of an exciton relative to the
$S_0$ ground state energy.
An intra-molecular transition accompanied with
electron spin flip, can be achieved by application of an
inhomogeneous perpendicular magnetic field
$\vec{B}_\perp = B_\perp(\vec{r})(\hat{x} \cos \omega_1 t
- \hat{y}\sin \omega_1 t)$, assuming $\omega_1 > 0$.
The quasi-particle energy levels shown in Fig.~\ref{Fig0},
form a $4\times 4$ block-diagonal Hamiltonian, $H$.
It can be decomposed to two $2\times 2$ Hamiltonians $H_1$ and $H_2$,
i.e., $H=H_1 \oplus H_2$.
Here $H_1=H^{(0)}_1 + H^{\perp}_1$ is
the two level Hamiltonian with the basis $|n\downarrow\rangle$, and
$|m\uparrow\rangle$ that form the diagonal matrix
$H^{(0)}_1 = \varepsilon_{n\downarrow} |n\downarrow\rangle\langle n\downarrow|
+ \varepsilon_{m\uparrow} |m\uparrow\rangle\langle m\uparrow|$ with
$\varepsilon_{n\sigma}=\varepsilon_n-\sigma\frac{\hbar\gamma}{2}B$.
Here $\gamma=g\mu_B/\hbar$ is the electron gyromagnetic ratio, and
the HOMO-LUMO single particle energy gap is
$\hbar\omega_{01} = \varepsilon_{m\uparrow} - \varepsilon_{n\downarrow}
= (\varepsilon_m - \varepsilon_n) - \hbar\gamma B$.
Application of $\vec{B}_\perp(\vec{r},t)$
allows a direct transition from spin down to spin up (as shown in
Figure~\ref{Fig0}). 
The Hamiltonian that describes the coupling of the molecular orbitals
with in-plane magnetic field can be expressed as
$H^{\perp}_1 = -\frac{1}{2}\hbar\gamma B^{nm}_\perp[
\exp (i\omega_1 t) |n\downarrow\rangle\langle m\uparrow| +
\exp (-i\omega_1 t)|m\uparrow\rangle\langle n\downarrow|]$.
Here $B^{nm}_\perp = \langle n| B_\perp(\vec{r}) | m \rangle$ is the
off-diagonal matrix element of inhomogeneous in-plane magnetic field
that allows mixing of HOMO and LUMO.
Finally
\begin{eqnarray}
H_1 &=& \left(\begin{array}{cc}
\varepsilon_{n\downarrow} & \Delta \exp (i\omega_1 t)   \\
\Delta^* \exp (-i\omega_1 t) &  \varepsilon_{m\uparrow}
\end{array}\right),
\end{eqnarray}
where $\Delta = -\hbar\gamma B^{nm}_\perp/2$.
Similarly
\begin{eqnarray}
H_2 &=& \left(\begin{array}{cc}
\varepsilon_{n\uparrow} & \Delta \exp (-i\omega_1 t)   \\
\Delta^* \exp (i\omega_1 t) &  \varepsilon_{m\downarrow}
\end{array}\right).
\end{eqnarray}
In $H_2$ the quasi particle energy gap is
$\hbar\omega_{02} = \varepsilon_{m\downarrow} - \varepsilon_{n\uparrow}
= (\varepsilon_m - \varepsilon_n) + \hbar\gamma B$.
Note that the oscillating in-plane magnetic field in $H_1$
provides the spin-flip resonance with the transition probability given by
$P_{n\downarrow\rightarrow m\uparrow} = (|\Delta/\hbar|^2/\Omega^2_R)
\sin^2(\Omega_R t/2)$.
Here $\Omega_R=\sqrt{(\omega_{01} - \omega_1)^2 + |\Delta/\hbar|^2}$.
Note that $\Omega_R=\Delta/\hbar$ is
the corresponding Rabi frequency at the resonance
$\omega_{01}=\omega_1$ where
the optical HOMO-LUMO transition accompany with the spin flip-flop
shows the highest probability.
In the last equality we ignored electron-hole Coulomb attraction.
Inclusion of $V$ is expected to shift the predicted resonance peak to
$\omega_1 = \Delta_X/\hbar$.
In contrast, because of negative sign of $\omega_1$ in
$H_2$, the oscillating in-plane magnetic field does not make a
resonance between opposite spins in HOMO and LUMO.
Therefore the spin flip transition induced by
$B_\perp$ occurs only between
$|n\downarrow\rangle$ and $|m\uparrow\rangle$, in accordance with
angular momentum conservation law.
The transition time for spin singlet to triplet of a $\pi$-pulse can be calculated by $\tau_p=(2\pi/g)(mc^2/e{\cal E}_\perp)/c$ where ${\cal E}_\perp=c \sqrt{\langle B^2_\perp\rangle}$.
The steady state population of spin-triplet state $n_T$ can be calculated by $n_T = n_0 \tau_r/(\tau_r+\tau_p)$ where $n_0$ is the density of the DNA-base in solution or on the surface-density of DNA-bases on the chromatin in nucleus of the cell accessible to the solvent (free radicals).
As $\tau_r>>\tau_p$, the optically pumped spin-triplet state quickly saturates.
A laser field with intensity of 1W/mm$^2$ is technologically achievable that gives $\tau_p\approx 10^{-7}$s.

\begin{figure}
\begin{center}
\includegraphics[width=1.0 \linewidth]{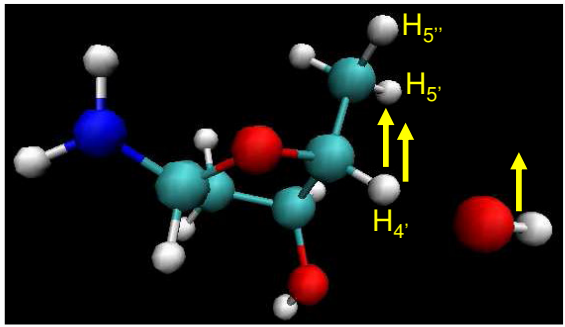}
\noindent
\caption{
Initial state of deoxyribose molecule in the presence of
OH free radical.
Schematically shown the injection of photo-generated electrons
in deoxyribose with spin polarization (shown by arrows)
along the direction of circularly polarized light and external
magnetic field.
The net magnetic force between two parallel magnetic moments
localized in OH and deoxyribose is repulsive.
This is similar to two separated magnetic moments which
interact like Heisenberg antiferromagnetic exchange coupling.
}
\label{Fig1a}
\end{center}
\end{figure}

\begin{figure}
\begin{center}
\includegraphics[width=0.8\linewidth]{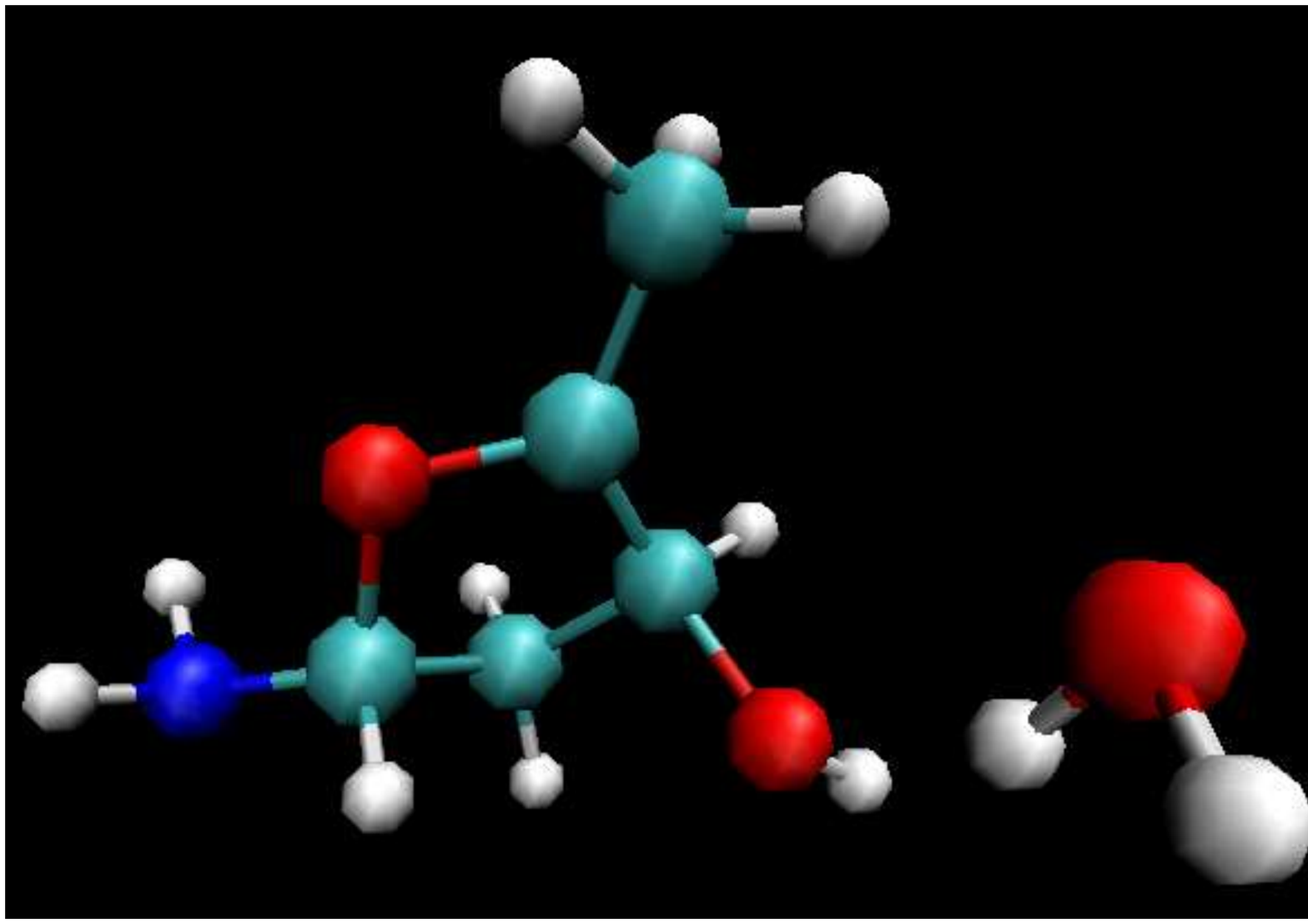}
\noindent
\caption{
The state of de-hydrogenated deoxyribose by OH free
radical at $t=0.15$ ps (a few time-steps after initial hydrogen abstraction).
In this simulation H$_{4'}$ on deoxyribose is abstracted.
The polarization state of the system is spin doublet ($S=1/2$).
}
\label{Fig1b}
\end{center}
\end{figure}

\begin{figure}
\begin{center}
\includegraphics[width=1.0\linewidth]{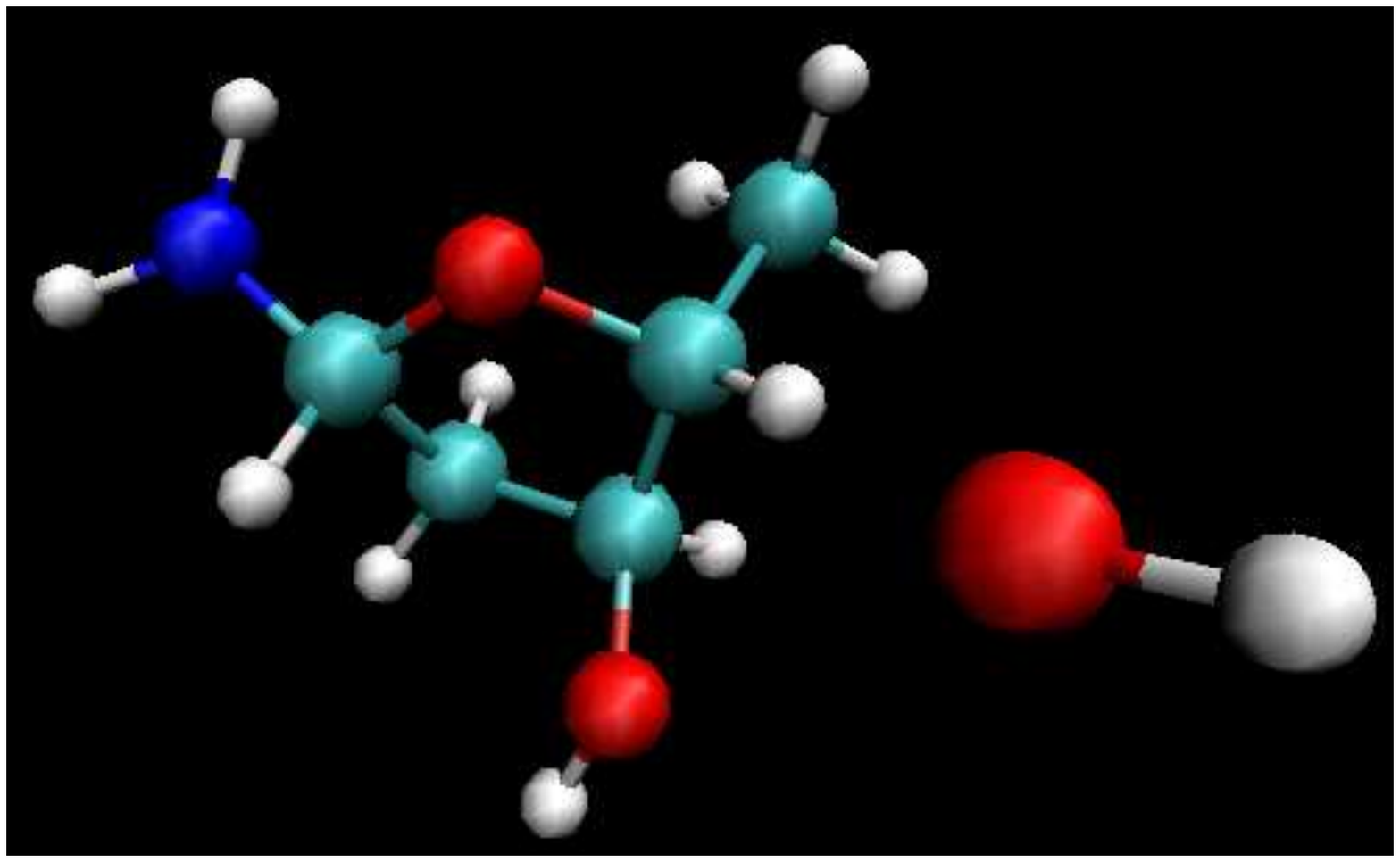}
\noindent
\caption{
The state of radio-resistive deoxyribose at $t=0.6$ ps.
The polarization state of the system is spin quartet ($S=3/2$) induced by
circularly polarized light in the presence of weak magnetic field.
Due to injected  localized and polarized photo-electrons, the
dehydrogenation of DNA-backbone does not occur.
}
\label{Fig1c}
\end{center}
\end{figure}

\begin{figure}
\begin{center}
\includegraphics[width=1.0\linewidth]{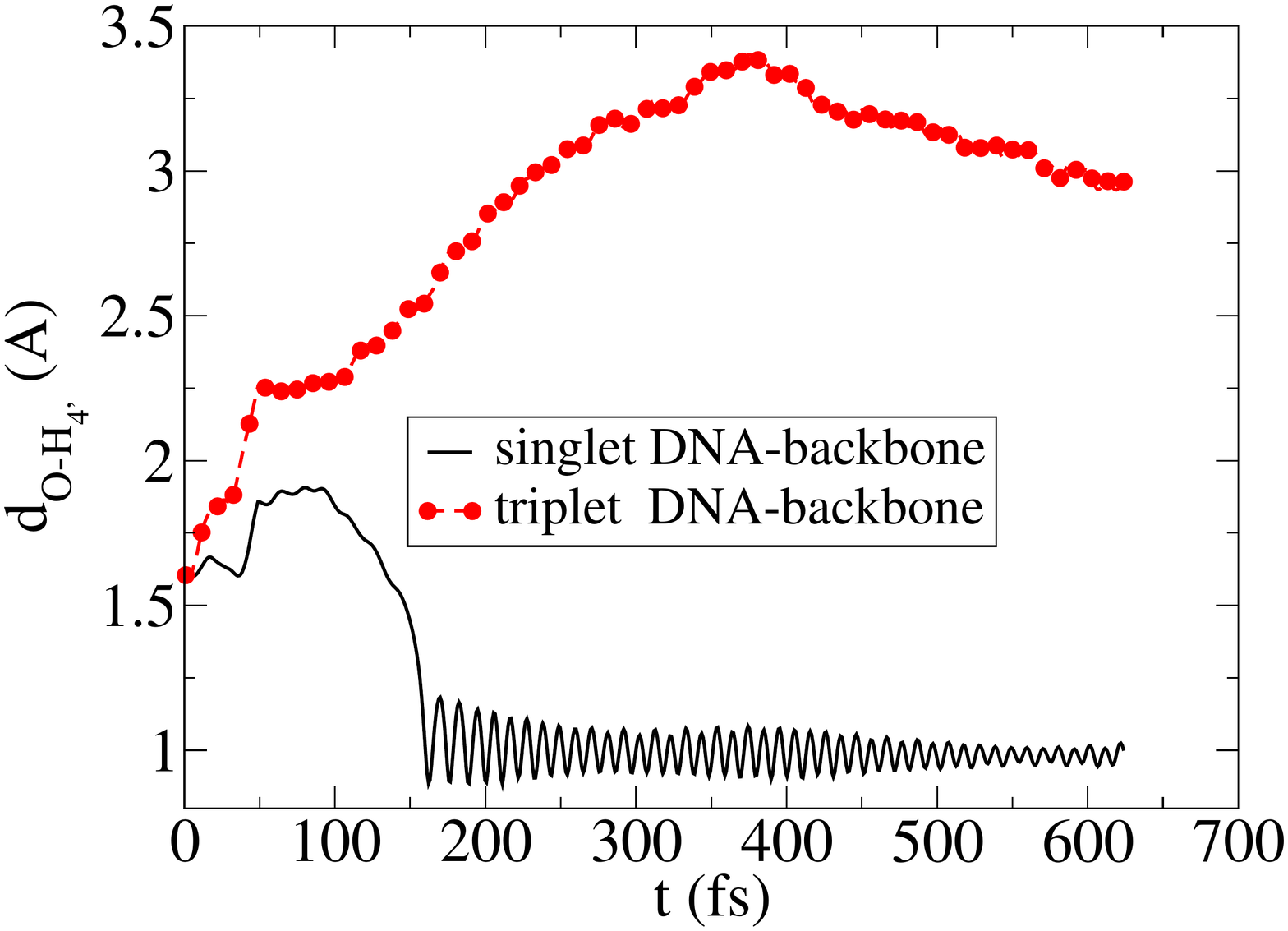}
\noindent
\caption{
Evolution of the separation of
oxygen atom in OH radical with respect to H$_{4'}$ in
deoxyribose. The dependence on the spin multiplicity is seen.
The hydrogen abstraction occurs approximately at $t=150$ fs in the spin
singlet state of deoxyribose.
The process of initial damage to DNA-backbone is blocked in deoxyribose
due to excessive spin with direction parallel to the spin polarization of
OH free radical.
}
\label{Fig2}
\end{center}
\end{figure}

\begin{figure}
\begin{center}
\includegraphics[width=1.0\linewidth]{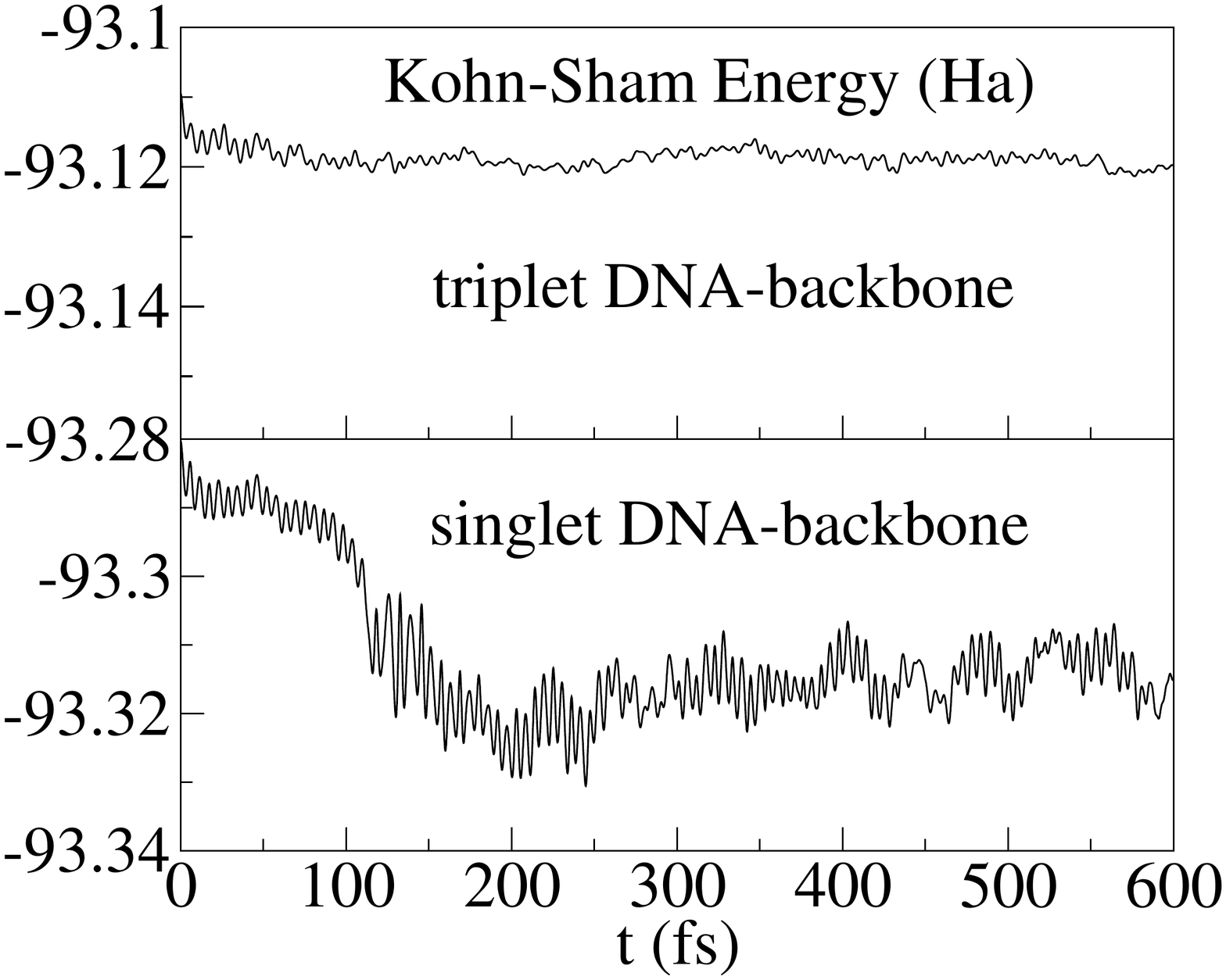}
\noindent
\caption{
Kohn-Sham energy as a function of time and spin-multiplicity of
deoxyribose.
An almost flat Kohn-Sham energy of spin triplet state of deoxyribose (top)
is indication of the repulsive interaction between magnetic moments of
OH free radical and $S=1$ exciton due to spin-blockade effect.
In contrast, a drop in Kohn-Sham energy of spin singlet state of
deoxyribose (bottom) indicates that the dehydrogenation of H$_{4'}$
occurs approximately at $t=150$ fs.
}
\label{Fig3}
\end{center}
\end{figure}

Figure~\ref{Fig1a} schematically shows optically excited electron-hole pair (exciton) in the DNA-backbone in the presence of spin-polarized OH free radical.
Because the absorption of a photon alters the
local electronic state of a DNA-molecule,
we confine our simulation to a particular segment, e.g., only a part of
DNA where the injected exciton
is presumably localized and the optical transition takes place,
e.g., close to the site of attack.
Our DNA-backbone model is similar to the deoxyribose residue
used in other {\em ab-initio} calculation such as Ref.
\cite{Miaskiewicz1994:JACS}
where the system of interest consists of a DNA nucleobase modeled by an
amino group attached to the deoxyribose in the presence of the OH-radical
as shown in Figure \ref{Fig1a}.
We selectively choose the site of OH-radical attack in the vicinity of
H$_{5'}$, H$_{5''}$, and H$_{4'}$ of deoxyribose residue in the back-bone of
DNA because of the accessibility of these sites to the solvent
\cite{Balasubramanian1998:PNAS,Pogozelski1998:CR,Tullius2005:COCB}.
Following Ref.~\cite{Abolfath2009:JPC}
we identify the dehydrogenation of the deoxyribose
as a function of spin multiplicity.
The ground and excited states of the deoxyribose correspond to
spin singlet ($S=0$), and spin triplet ($S=1$) states.
As the life time of both spin-polarized OH radical and spin-triplet exciton in
DNA-backbone is much longer than the simulation time, we perform CPMD~\cite{CPMD}
calculation under restricted spin configuration.
We choose two classes of spin-restricted calculations, as
the total spin along the quantum axis
is subjected to the constraints $S_z=1/2$, and $3/2$,
corresponding to doublet and quartet spin configurations.
In both calculations the initial distance between OH-radical and DNA-backbone
is considered to be approximately 1.5 $A$.

The initial and final states of the molecules are shown in
Figures~\ref{Fig1a}-\ref{Fig1c}.
The final configurations of the molecules have been obtained after 0.6 ps
where the rearrangement of the atomic coordinates have been
deduced from a dynamical trajectory calculated by CPMD.
According to our results, a dehydrogenation
of the deoxyribose takes place around 0.15 ps
for a system with $S_z=1/2$ (total spin-doublet) as shown in Figure \ref{Fig1b}.
This process leads to the formation of a water molecule, and H$_{4'}$
turns out to be the site of Hydrogen abstraction in this simulation.
In contrast, as shown in Figure \ref{Fig1c},
in the quartet spin configuration the repulsive exchange
interaction
blocks the exchange of hydrogen
and hence the chemical reaction.
To systematically check the convergence of the results, we
increased the size of the DNA-molecule by adding phosphates and
nucleotides and constructed a double strand DNA.
We used a quantum mechanical (QM) approach to simulate the Hydrogen abstraction in a small segment of DNA molecule and molecular mechanics (MM) approach to simulate the vibrational modes of the rest of molecule~\cite{Gervasio2004:CEJ}, and found no influence on the spin-blocking effect.
The details of this study will be presented elsewhere.
To estimate the energy needed for the polarization of the deoxyribose
in the absence of OH-radicals,
we calculated the energy of the ground and excited states of the gas-phase
deoxyribose in spin singlet and triplet multiplicities with
the spin singlet-triplet energy gap
$\Delta_X \approx 2.38$ eV.
This provides an estimate for the frequency of the
optical gap shown in Figure~\ref{Fig0}, which is within the range of the
visible spectrum of the electromagnetic waves, $\lambda=521$ nm,
corresponding to green light.
To estimate the stored magnetic energy due to the optical injection of spin,
we calculated the energy of the gas-phase deoxyribose
in the presence of one OH-free radical with spin doublet and
quartet multiplicities.
For the molecules shown in Figure \ref{Fig1a}, we find the energy gap
${\tilde\Delta}_X \equiv E_{\rm quartet} - E_{\rm doublet} \approx 3.05$ eV.
Here the excessive magnetic energy which originated from
spin-spin repulsive interactions (which resemble the anti-ferromagnetic
exchange interaction in the Heisenberg model) can be deduced to be
${\tilde\Delta}_X - \Delta_X \approx 0.67$ eV.
This energy can be interpreted as the excessive energy barrier
due to the alignment of the spins in deoxyribose and OH, and is
the source of the magnetic repulsive force which makes the diffusion of OH
toward DNA-backbone less likely.
This is in agreement with the results obtained from CPMD.
The distance between the oxygen atom in OH radical and H$_{4'}$ in
deoxyribose as a function of time and the spin multiplicity of deoxyribose
is shown in Figure~\ref{Fig2}.
A drop in Kohn-Sham energy seen in spin singlet state that
is the indication of deoxyribose dehydrogenation, disappears in case of spin triplet state.
Accordingly dehydrogenation of H$_{4'}$ occurs approximately at $t=150$ fs.
The corresponding Kohn-Sham energies are shown in Figure~\ref{Fig3}.
In spin triplet state of deoxyribose the spin-blockade effect is strong enough
that it repels OH free radical.

In conclusion, we have examined spin-blockade mechanism
to propose coherent control of DNA-OH interaction, in particular from
the initial damage induced by OH free radicals.
To achieve an efficient direct pumping of triplet states from the singlet ground state we illustrated that the photo-excitation pumping rate can be tuned to be faster than the reported phosphorescence life-time, however, an indirect transition via $S_0 \rightarrow$ $^1\pi\pi^* \rightarrow$ $^3\pi\pi^*$ is
an alternative optical pumping pathway that can be used.
The model calculation was performed using ab-initio Car-Parrinello molecular
dynamics of a deoxyribose residue attached to amino-group in the presence of
OH free radicals.
We provided a detailed description of an experimental set up for
spin-blockade detection.

\end{document}